# ON THE STRUCTURAL AND OPTICAL PROPERTIES OF SPUTTERED HYDROGENATED AMORPHOUS SILICON THIN FILMS

**A. BARHDADI [1*] and M. CHAFIK EL IDRISSI [2]**

[1] *Laboratoire de Physique des Semiconducteurs et de l'Energie Solaire (P.S.E.S.),
Ecole Normale Supérieure de Rabat, P.O.Box: 5118, Rabat 10000, Morocco
and
The Abdus Salam International Centre for Theoretical Physics, Trieste, Italy.*

[2] *Laboratoire de Physique des Surfaces et Interfaces (L.P.S.I.),
Faculty of Sciences, Ibn Tofail University, P.O.Box: 133, Kénitra 14000, Morocco*

___________________________________________________________________________

[*] Corresponding author,
Regular Associate of the Abdus Salam ICTP
Phone:    (212) 37 75 12 29 or 37 75 22 61 or 64 93 68 15
Fax:      (212) 37 75 00 47,
E-mail:   abdelbar@fsr.ac.ma

# Abstract

The present work is essentially focused on the study of optical and structural properties of hydrogenated amorphous silicon thin films (a-Si:H) prepared by radio-frequency cathodic sputtering. We examine separately the influence of hydrogen partial pressure during film deposition, and the effect of post-deposition thermal annealings on the main optical characteristics of the layers such as refraction index, optical gap and Urbach energy. Using the grazing X-rays reflectometry technique, thin film structural properties are examined immediately after films deposition as well as after surface oxidation or annealing. We show that low hydrogen pressures allow a saturation of dangling bonds in the layers, while high doses lead to the creation of new defects. We show also that thermal annealing under moderate temperatures improves the structural quality of the deposited layers. For the films examined just after deposition, the role of hydrogen appears in the increase of their density. For those analysed after a short stay in the ambient, hydrogen plays a protective role against the oxidation of their surfaces. This role disappears for a long time stay in the ambient.

# I- INTRODUCTION

Thin film technologies hold considerable promise for a substantial reduction of the manufacturing costs of solar cells due to the reduction of material costs and the deposition on large area substrates [1 - 5]. The thin film technologies for photovoltaic applications include a-Si:H alloys, CdTe, Cu(In,Ga)Se$_2$ (CIGS), poly-Si, μc-Si/Poly-Si and dye/TiO$_2$. The most advanced technologies are a-Si-alloys, CdTe and CIGS [1, 2]. Solar cells and a range of other electronic devices technologies based on hydrogenated amorphous silicon (a-Si:H) have matured considerably over the last two decades. Since the first a-Si:H solar cell was made by Carlson et al. [6] the technology has improved tremendously, leading to reported initial conversion efficiencies exceeding 15% [7].

Solar cells based on a-Si:H offer many specific advantages. They are fabricated from extremely abundant raw materials and involve almost no ecological risk during manufacturing, operation, and disposal. With comparison to crystalline silicon, the a-Si:H material used in the fabrication is distinguished by a large optical gap in the required range for optimal performance [8]. It presents also a high optical absorption within the maximum of solar spectrum. So, a large part of solar energy is absorbed in a small thickness of the material allowing making structures in the form of very thin layers. Indeed, a thickness of 1 μm material suffices to absorb the solar radiation efficiently. Also, a-Si:H does not considerably suffer light induced degradation (Staebler-Wronski effect) [9]. Moreover, deposition of a-Si:H is much faster than crystalline silicon growth and can be carried out over much larger areas. Low process temperatures facilitate the use of a variety of low cost substrate materials such as float glass, metal or plastic foils [10-12]. All these characteristics are of great importance making a-Si:H a more attractive material for the elaboration of cheap photocells with good photovoltaic parameters.

Several methods have been used to deposit amorphous silicon from silane or other silicon carrier gases [13, 14]. These include: chemical vapour deposition (CVD), direct current (DC) and radio frequency (RF) glow discharge (plasma enhanced chemical vapour deposition), microwave glow discharge, electron-cyclotron resonance glow discharge, remote plasma-assisted CVD, controlled plasma magnetron (CPM) glow discharge, photolytic decomposition (photo-CVD), sputtering, cluster beam evaporation and hot-wire decomposition. Of these, RF and DC glow discharge depositions are the most common and are those used by industry. The optimum a-Si:H material has traditionally been produced at a substrate temperature of around 250 °C and contains about 10 at.% H. Some gas mixtures used to deposit a-Si:H contain SiH$_4$, Si$_2$H$_6$, SiF$_4$ and H$_2$, while others may use only SiH$_4$. The most recent route to obtain higher quality films is to use H-dilution of the feedstock gases. This improves the initial film quality in some systems and leads to less light induced degradation in solar cells.

In the present work, we prepare a-Si:H thin films by radio frequency cathodic sputtering technique and we study how optical characteristics of these films change with increasing hydrogen pressure during the deposition stage as well as with classical post-deposition annealing. Using the grazing x-rays reflectometry technique, we also characterise some of the structural properties of a-Si:H very thin layers immediately after deposition as well as after surface oxidation or annealing.

## II- EXPERIMENTAL DETAILS

### II -1- Sample preparation

Starting from a silicon polycrystalline target under reactive ions of argon and hydrogen plasma, a-Si:H thin films of about 0.5 µm thick, and 1 cm$^2$ area are deposited by radio frequency cathodic sputtering technique [15]. The experimental set-up we have used is a full computerised system, model SCM 451, from Alcatel. It consists of a sputtering enclosure which allows depositions at different temperatures, a pumping system allowing reaching a vacuum of nearly $10^{-7}$ mbar, and a radio-frequency source with a 0 – 500 W power range. The substrates we have used are made of corning glass C 7059. Before any deposition, both the sputtering enclosure and substrates are submitted to an appropriate cleaning to prevent the eventual contamination of silicon films. Also the target undergoes an ionic etching during approximately 15 min.

A-Si:H thin films have been deposited by sputtering the polycrystalline silicon target under a constant total pressure of about $2.10^{-2}$ mbar and 250W radio-frequency power. In our operating conditions, the substrate is maintained at the ambient temperature (around 25 °C) which, under the effect of the plasma formed inside the enclosure, can reach 100 °C [16]. The partial pressure of hydrogen gas introduced into the enclosure ($P_H$) being among of the main parameters that we intent to study the effect, it was then ranged from 0 to 75 % of the total pressure.

To calibrate well our experimental set-up, many a-Si:H layers have been first deposited at various $P_H$ during a fixed time $\Delta t$. Next, for each layer the corresponding transmission spectrum was measured and then the layer thickness $\Delta e$ was deduced. After, the deposition rate $V_d$ was calculated for each $P_H$ using the simple ratio: $V_d = \Delta e/\Delta t$. Thus, to obtain a thin film with a thickness $\Delta e$, we just have to fix the deposition time $\Delta t$ calculated from dividing $\Delta e$ by the deposition rate $V_d$ corresponding to the chosen $P_H$ value during the film deposition. Figure 1 shows the evolution of $V_d$ with the partial pressure of hydrogen $P_H$. We see that $V_d$ decreases exponentially as $P_H$ increases but the variation is less significant beyond $P_H = 8.10^{-3}$ mbar leading to a quasi-saturation of $V_d$ for $P_H \geq 10^{-2}$ mbar.

Thin films obtained are stoechiometric with a good adherence. They are supposed homogeneous and having uniform thickness. A detailed study of thickness mapping performed on thin films obtained by a deposition system very similar to ours is described in the literature [17, 18]. It consists to calculate the thickness at several points of layers of 4 to 5 x 2 cm$^2$ surface. At the edges of layers, calculated thickness was found to be weaker than that calculated in the centre. However, for layers of about 1 cm$^2$ surface, good thickness homogeneity was observed. This is why we often used substrates with a small surface. Larger ones have been sometimes used but equipped with masks. Also substrates were always placed at an optimal distance from the target (6 cm) and well centred on its axis.

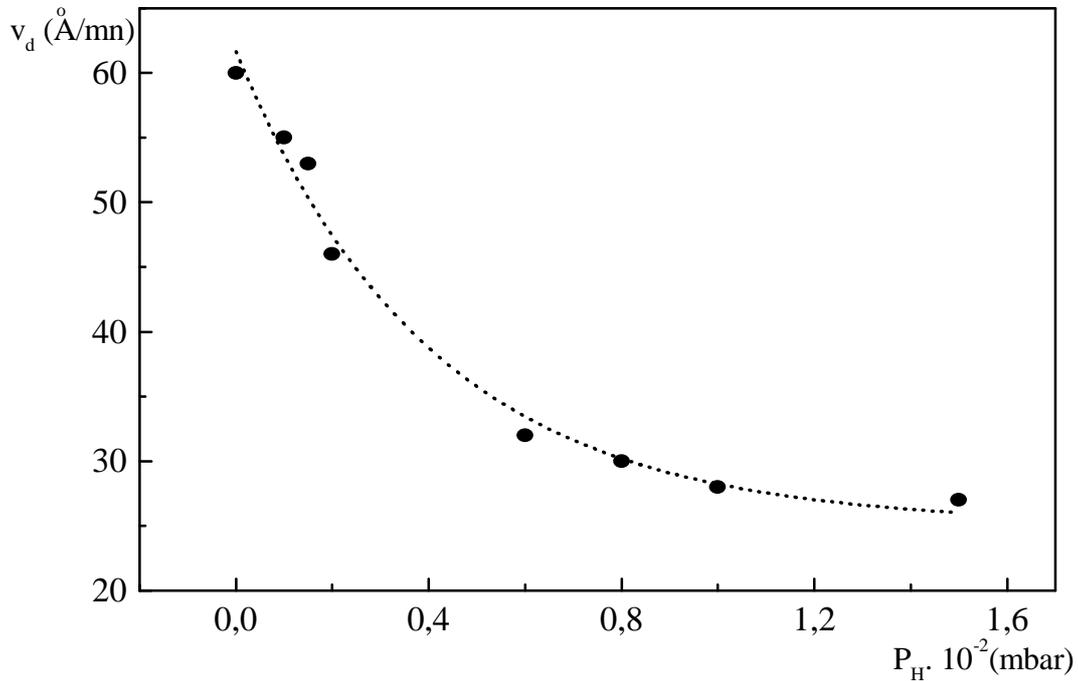

**FIGURE 1**
*Evolution of the deposition rate $V_d$ according to the partial pressure of hydrogene $P_H$.*

## II -2- Experimental techniques and analyses

Structural properties of deposited films have been examined by the grazing X-rays reflectivity technique [19]. This well-known technique is of great interest in surface science, since it allows the structure of the uppermost layers of a material to be probed. It consists in sending on the sample a monochromatic parallel beam of X-rays in the grazing incidence condition and recording the evolution of the reflected beam intensity I as a function of the variations of the incidence angle $\alpha$. The analysis of the experimental curves I($\alpha$) obtained allows the determination and/or the calculation of thickness, refractive index, roughness and both mass and electronic densities of very thin films while noise and other parasitic effects resulting from the films supports are minimised [19]. Some details of the technique related calculations were described elsewhere [20]. To determine parameters such as thickness, mass and electronic densities, the method of calculation depends closely on the how many layers composing the considered film. In the literature, we found several calculations methods [21, 22]. Those we have used to characterise our mono-layer and bi-layer deposited thin films are sufficiently explained in our previous work [20].

Optical characteristics of deposited layers were determined by using a Shimadzu UV-3101 PC spectrophotometer operating with a double beam in a wide spectral range: from ultraviolet to near infrared light. Transmission measurements were taken by means of the differential method [23] which consists in determining the transmission of the layer together with its substrate by comparison to another identical virgin substrate used as reference. Under these conditions, the influence of the substrate on the transmitted light is practically negligible [23].

## III - RESULTS AND DISCUSSIONS

Some of deposited thin films were characterised immediately after deposition without undergoing any particular processing. Others were examined after being oxidised in the ambient or thermally annealed.

### III -1- Measurements performed just after sample deposition

These measurements were performed on a-Si:H thin films deposited under various partial pressure of hydrogen gas $P_H$ ranging from 0 to $1.5 \times 10^{-2}$ mbar (i.e. 0 to 75% of the total pressure). In order to ensure of the structural quality of as-deposited layers, X-rays diffraction measurements have been systematically performed on all samples [24]. The results gave spectra similar to that presented on figure 2 confirming the amorphous character of our layers.

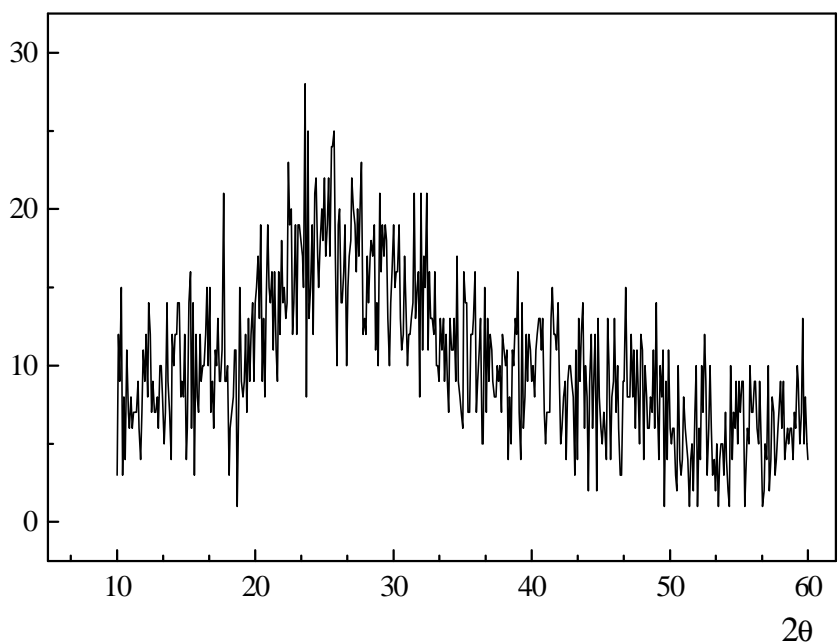

**FIGURE 2**
*X-rays diffraction spectrum of a-Si:H thin film deposited under $P_H = 1.5 \times 10^{-2}$ mbar.*

Figure 3 shows a typical transmission spectrum (normalized to the substrate) obtained for an a-Si:H thin film deposited under a $P_H$ of $1.5 \times 10^{-2}$ mbar. All the other transmission spectra obtained for the different a-Si:H samples deposited under various $P_H$ present practically the same form [25, 26]. They all present two clearly distinct slopes [27]. The first one, in which interference fringes (oscillations) are seen, is commonly named transparency zone or zone of weak absorption. In the second one, which is well known as zone of high absorption, the signal is strongly reduced.

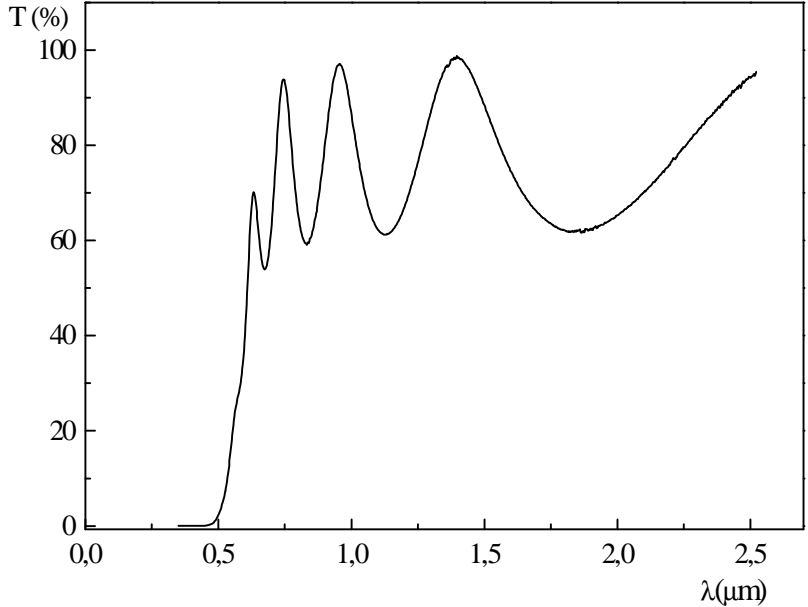

**FIGURE 3**

*Transmission spectra, normalised to the substrate, obtained for an a-Si:H thin film deposited under a partial pressure of hydrogen $P_H$ of 1.5 $10^{-2}$ mbar*

By exploiting the different transmission spectra corresponding to the various hydrogen partial pressures, and using the mathematics expressions reported in the reference [28], we were able to determine the main optical characteristics of our samples: layer thickness, surface roughness, refraction index, optical gap ($E_g$) and energy of Urbach ($E_u$) [29]. It should be interesting to remember that the latter draws light on the density of the energy states localised in the tail of the valence band, which are generally attributed to the structural disorder in the material [30].

In figure 4, we reproduce the dependence of deposited thin films refraction index (n) with the wavelength ($\lambda$) for all $P_H$ used. The values of n were adjusted using the dispersion law (equation 1) of Sellmeiere [31]. It is clear that, when $P_H$ increases, $n(\lambda)$ curve shifts towards higher n values. This is in good agreement with the results published by Swanpoel and Swart [32].

$$n^2(\lambda) = n_\infty^2 + \frac{b^2}{\lambda^2 - \lambda_0^2} \qquad (1)$$

where $n_\infty$ is the refraction index obtained by extrapolation towards the infinite. b and $\lambda_0$ are constants determined by the $n(\lambda)$ curve fitting.

The increase in n is due to a thickening of the layers by the presence of hydrogen. Indeed, hydrogen atoms densify the layers through the elimination of dangling bonds existing at very high content in amorphous materials [32, 34], and hydrogen molecules $H_2$ fill the vacant sites present in these layers [35].

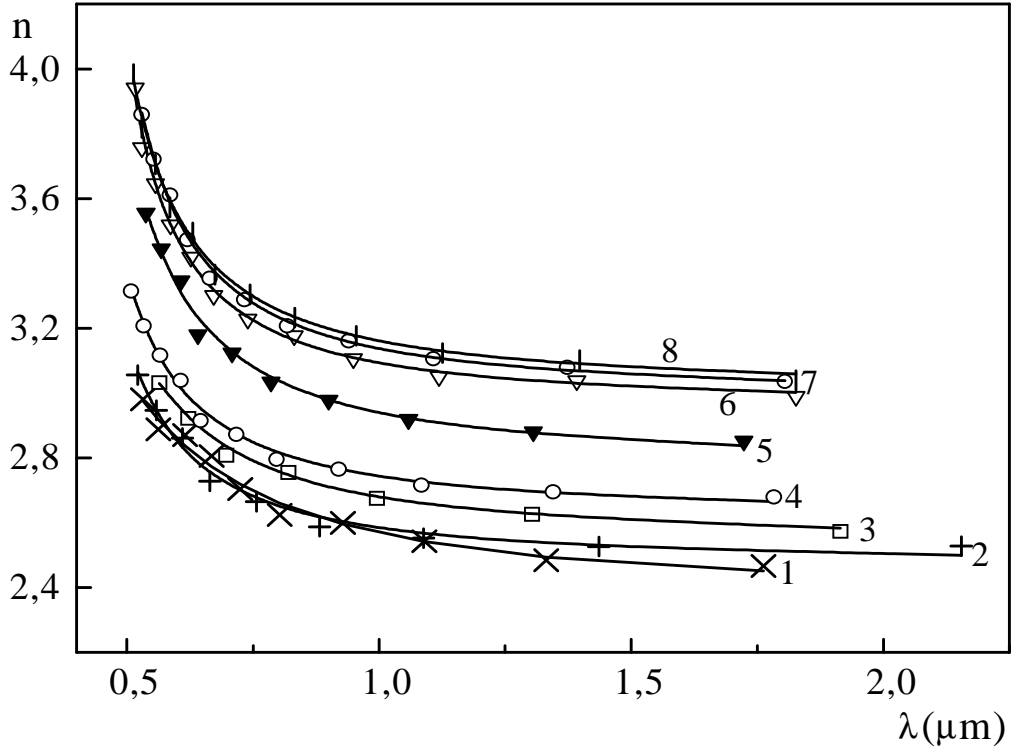

**FIGURE 4**

*Experimental curves showing the evolution of the refraction index n of the deposited layers as function of the wavelength λ for various $P_H$. Experimental data points were fitted to the Sellmeiere formula [31].*
*1) $P_H = 0$; 2) $P_H = 2.10^{-3}$ mbar; 3) $P_H = 3.10^{-3}$ mbar; 4) $P_H = 4.10^{-3}$ mbar; 5) $P_H = 6.10^{-3}$ mbar; 6) $P_H = 8.10^{-3}$ mbar; 7) $P_H = 10^{-2}$ mbar; 8) $P_H = 1,5.10^{-2}$ mbar.*

Figure 5 shows the evolution of $E_g$ and that of $E_u$ as function of $P_H$ pressure. $E_g$ values were determined from Tauc formula [36] expressed by equation 2 using figure 6 plotting, and those of $E_u$ from Urbach law [29] formulated by equation 3 using figure 7 graphing.

$$(\alpha h\nu)^{1/2} = B (h\nu - E_g) \qquad (2)$$

$$\alpha(h\nu) = \alpha_0(h\nu) \exp\left(\frac{h\nu - h\nu_0}{E_u}\right) \qquad (3)$$

α is the absorption coefficient, B is a coefficient of proportionality, $\alpha_0$ and $h\nu_0$ are constants depending on depositing conditions of thin films.

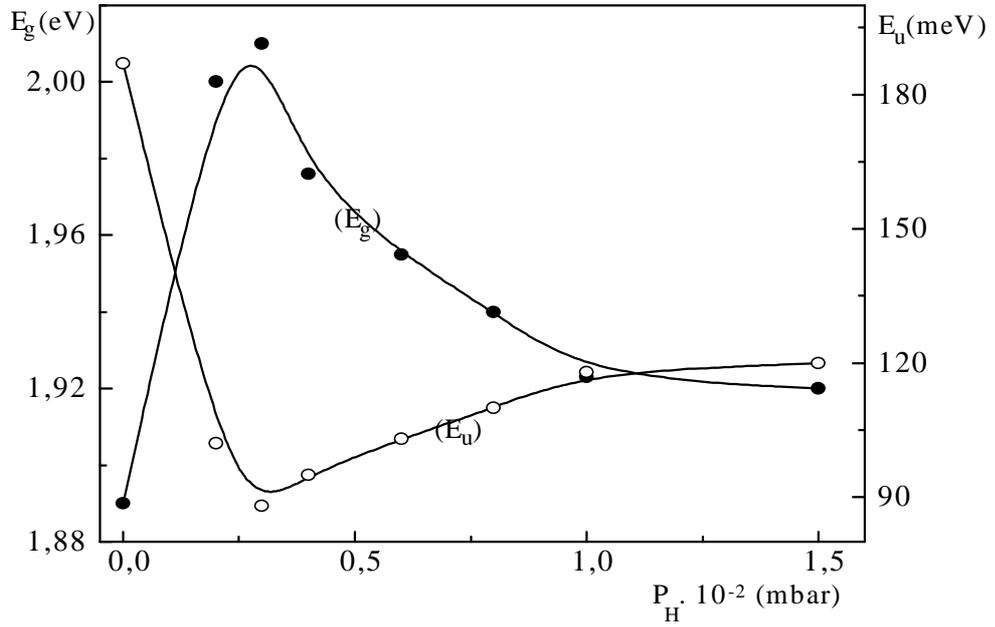

**FIGURE 5**
*Evolution of the optical gap ($E_g$) and of the Urbach energy ($E_u$) measured on deposited layers as function of $P_H$. $E_g$ was determined from the Tauc formula [36] and $E_u$ from Urbach law [29]. We show that both $E_g$ and $E_u$ exhibit strictly opposed evolution trends.*

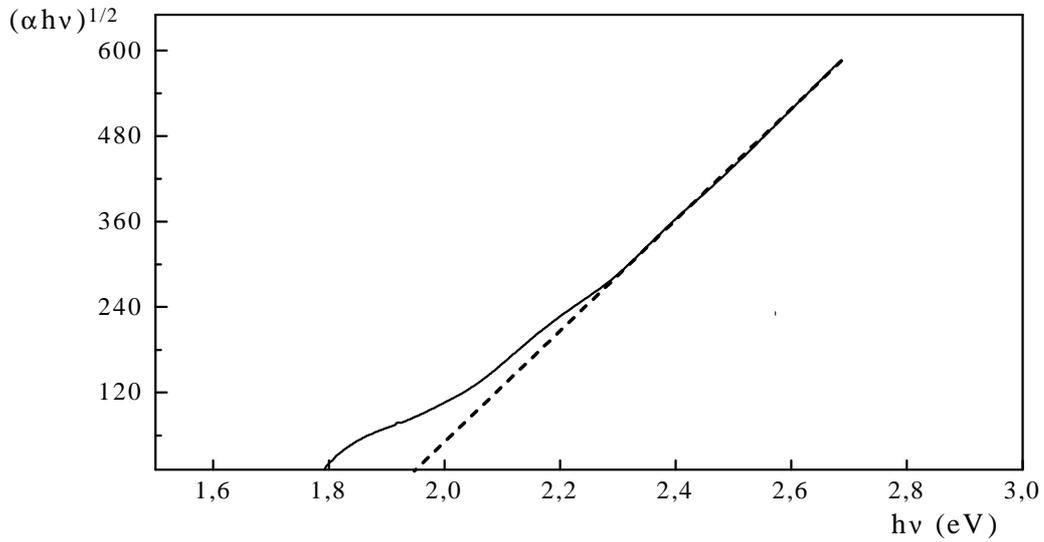

**FIGURE 6**
*Determination of $E_g$ from Tauc formula [36]. The curve in dotted line is the extrapolation of the Tauc formula at high energies.*

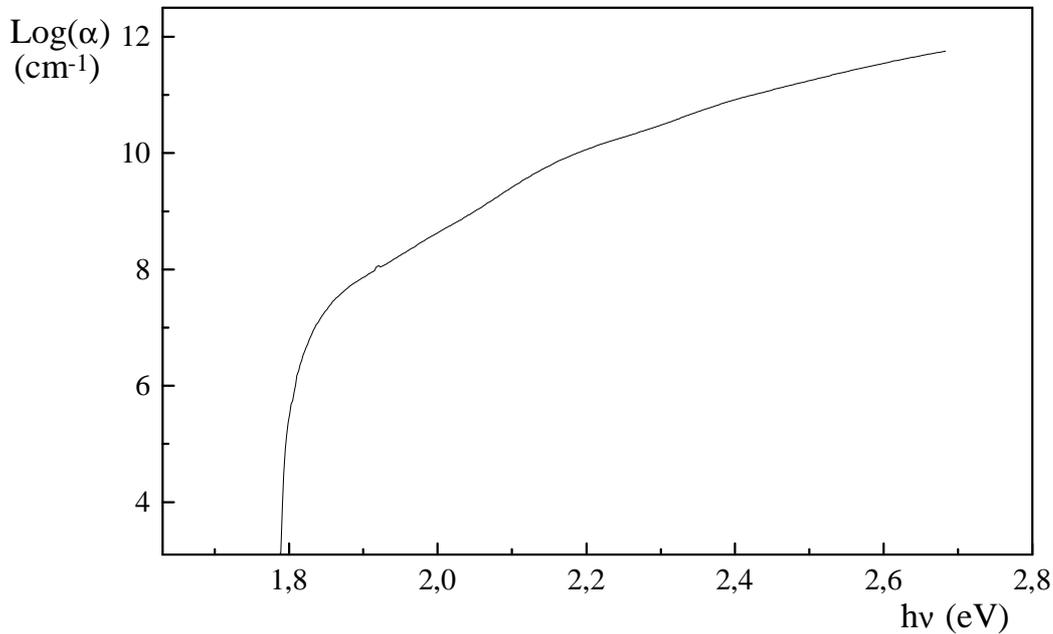

**FIGURE 7**

*Curve showing in semi-logarithmic scale the variation of the absorption coefficient according to the energy for the calculation of $E_u$ from Urbach law [29].*

We notice that $E_g$ and $E_u$ evolve into strictly opposing directions: when one increases the other decreases, and when the first culminates, the second reaches its minimum. This clearly confirms the very close relationship existing between these two parameters, because of the presence of a more or less important density of dangling bonds in the material.

For low $P_H$ values, we observe an important and fast increase of $E_g$ which quickly reaches its maximum at $P_H$ of $3.10^{-3}$ mbar. Beyond this threshold, $E_g$ starts decreasing exponentially before saturating at a value slightly higher than that of the control. Co-jointly, $E_u$ perfectly develops in the opposite direction. From these results, we can say that for low $P_H$, hydrogen incorporation saturates dangling bonds in the deposited layers. This leads to a clear improvement of $E_g$ and $E_u$. With a $P_H$ value of $3.10^{-3}$ mbar, the density of dangling bonds is minimal and, hence, the optical performances of the layers are optimal ($E_g$ culminates and $E_u$ is minimal). When $P_H$ becomes higher than its optimal value, excessive presence of hydrogen in the material generates new structural defects, the concentration and the complexity of which increase with the density of the incorporated hydrogen species. This has negative repercussions on the optical performances of the layers, and leads to a re-diminution of $E_g$, accompanied by an increase of $E_u$ as we can see on figure 5.

To look at the effect of hydrogen pressure on some of structural properties of sputtered a-Si:H thin films we have deposited, some measurements were made by grazing X-rays reflectometry technique [19]. In figure 8, we have plotted in semi-logarithmic scale the experimental curves giving the evolution of the reflected intensity I according to the incidence angle α for three samples A, B and C deposited under hydrogen partial pressure $P_H$ of 0, $10^{-3}$ and $2.10^{-3}$ mbar respectively, and characterized immediately after deposition.

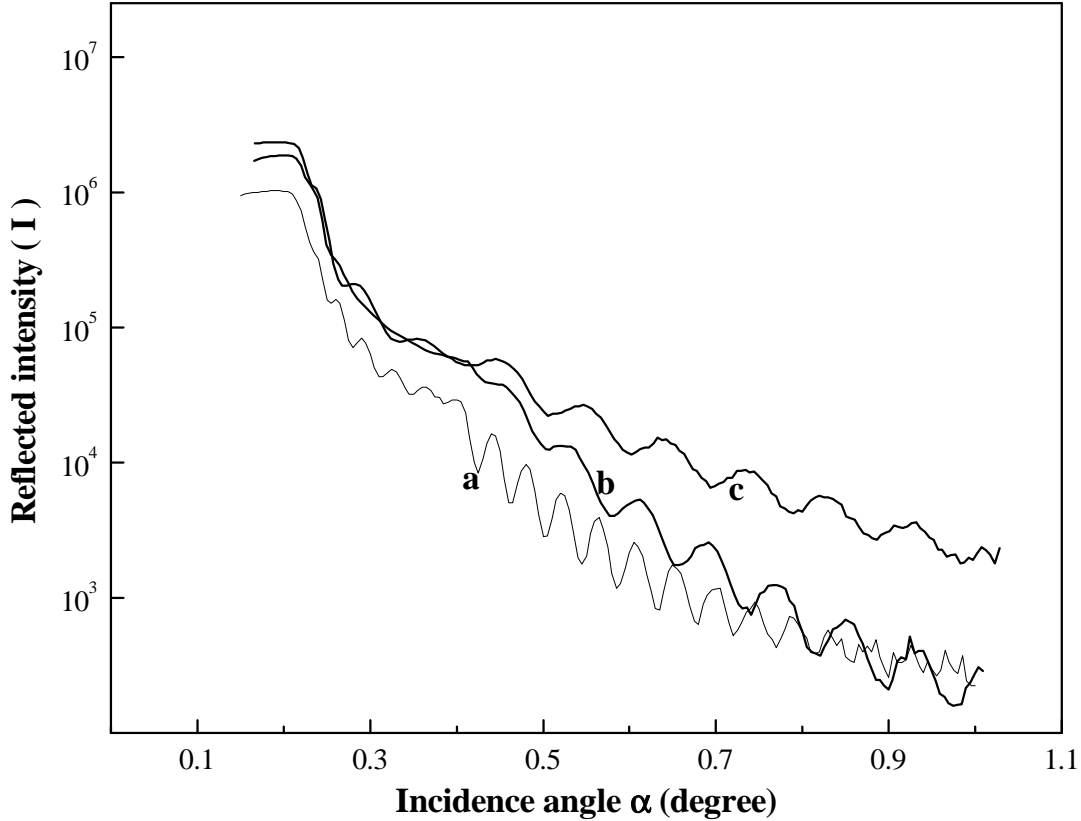

**FIGURE 8**

*Semi-logarithmic plotting of the reflected intensity I according to the incidence angle α for three samples A, B and C examined immediately after deposition.*
*Curve a is corresponding to sample A prepared without hydrogen incorporation ($P_H = 0$ mbar)*
*Curve b is corresponding to sample B prepared with hydrogen partial pressure $P_H = 10^{-3}$ mbar*
*Curve c is corresponding to sample C prepared with hydrogen partial pressure $P_H = 2.10^{-3}$ mbar*

Using the calculation method we have described previously [20], the linear fitting of the above experimental curves was performed by drawing $\sin^2(\alpha_p)$ versus $(p+1/2)^2$ where p is the fringe order and $\alpha_p$ its angular position [20]. The exploitation of the obtained fitting lines was made using a program based on the least squares method. In table I, we have presented the results deduced from this experimental curves processing. We are particularly interested in the results relating to the two principal parameters that are the thickness $\Delta e$ and the mass density $\rho$ of the deposited layers. We note that $\rho$ increases with the hydrogen partial pressure $P_H$, whereas $\Delta e$ evolves in the opposite direction. These results agree well with those obtained before from the optical measurements. The increase of ρ is explained by that observed on the refraction index n and the thickness decrease is due to the reduction of the deposition rate [24]. ρ increasing reflects the positive role played by hydrogen in the layers thickening resulting from an effective saturation of dangling bonds in the films.

| Samples | Partial pressure of H $P_H$ (mbar) | Thickness $\Delta e$ (Å) | Critical angle $\alpha_c$ (degree) | Electronic density $N_Z$ (e . cm$^{-3}$) | Mass density $\rho$ (g . cm$^{-3}$) |
|---|---|---|---|---|---|
| A | 0 | 958 | 0.2026 | 5.8649 10$^{23}$ | 1.9554 |
| B | 10$^{-3}$ | 522 | 0.2185 | 6.8193 10$^{23}$ | 2.2736 |
| C | 2.10$^{-3}$ | 470 | 0.2187 | 6.8327 10$^{23}$ | 2.2781 |
| D | 2.10$^{-3}$ | 445 | 0.2207 | 6.9557 10$^{23}$ | 2.3191 |

**TABLE 1**
*Parameters values determined from X-rays analyses performed on the samples A, B, C, and D. Sample D has been deposited under the same experimental conditions as sample C but before being measured by grazing x-rays reflectometry it has been thermally annealed for 30 min in a conventional furnace at a temperature of 500 °C. Comments on the corresponding obtained results are given below.*

**III -2- Measurements performed after thermal annealing**

After examining the effects of partial hydrogen pressure on the optical and structural characteristics of a-Si:H thin films immediately after their deposition, we focused on the changes that would affect these characteristics after thermal annealing. It is essential to bear in mind that stability of photovoltaic a-Si:H compounds depends on their normal functioning temperature and on the thermal solicitations to which they are sometimes compelled to be submitted. This is why many studies hold on this extremely important aspect of the problem [33, 37-40].

In order to conduct this part of work, we have prepared three a-Si:H samples under experimental conditions similar to those mentioned above, using a $P_H$ value of 1,5.10$^{-2}$ mbar. One of these samples is kept as a control. The other two ones were each submitted to a 45 min thermal annealing in a conventional furnace under a vacuum of 10$^{-6}$ mbar. Annealing temperature ($T_R$) was chosen to be the variable parameter of which we try to determine the effect. $T_R$ was fixed at 350 °C for the first annealing and at 600 °C for the second one.

Figure 9 shows the spectral scattering of the refraction index n for each of the three examined samples. We notice that thermal annealing induces a clear n increasing in the whole-explored spectral range. This increase is obviously slight for $T_R$ = 350 °C but, when $T_R$ reaches 600 °C, it becomes so important that, for high λ, n reaches typical values usually measured on silicon crystals [41]. In regard to this result, we can already say that thermal annealing at 600 °C allows a partial crystallisation of the deposited layer. Thus, we meet literature data which show that amorphous silicon crystallises at a critical temperature between 550 °C and 700 °C during annealing processing [33, 39].

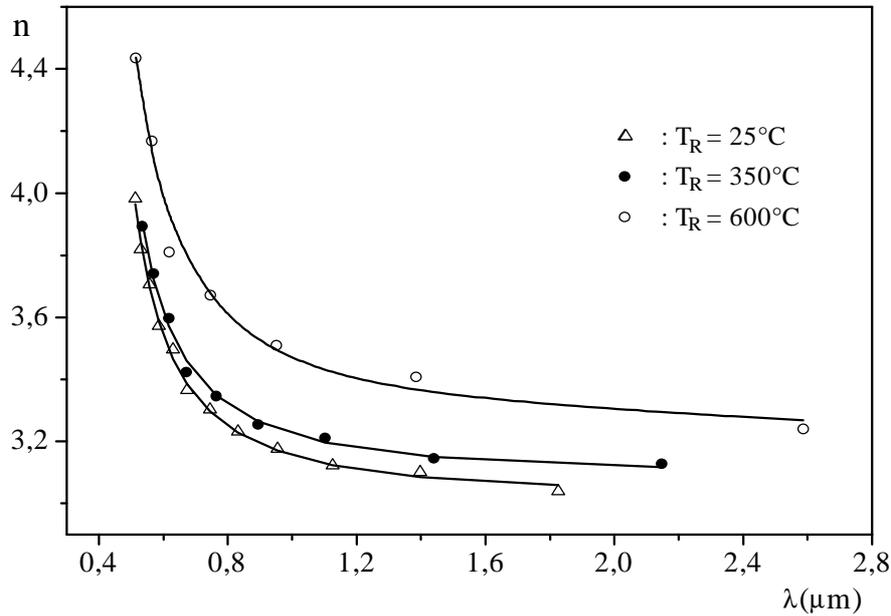

**FIGURE 9**

*Spectral scattering of the refraction index n and its evolution with the annealing temperature $T_R$*

In figure 10, the variations of $E_g$ and those of $E_u$ with the annealing temperature $T_R$ were both reported on the same graph. The first remark is the close relationship that links these two parameters making their evolution curves quite symmetric. Furthermore, these curves present an important similarity with those found in the literature [42]. Examination of these curves allows noting the following points:

i) For the reference sample (control), $E_g$ and $E_u$ present values practically equal to those measured above on the sample that was prepared in the same experimental conditions (figure 5). This proves that our results are reproducible. In other words, the structural quality of the deposited layers is practically influenced only by deposition experimental conditions which are entirely under control.

ii) The first annealing at 350 °C simultaneously leads to a decrease in $E_g$ and to a proportional increase in $E_u$. On the basic of literature data [42, 43], we attribute this to a partial out-diffusion of hydrogen resulting from a break of the less stable bonds of $SiH_2$ and $SiH_3$ that have been formed in the material during the deposition.

iii) After annealing at 600 °C, $E_g$ decreases considerably reaching a value of 1.57 eV. Simultaneously, $E_u$ increases to reach an average value of 138 meV. Logically, and in the same way as in the first annealing operated at 350 °C, these variations can also be attributed to the out diffusion of a higher quantity of hydrogen as a result of the break in the more stable bonds of SiH [42, 44]. Nevertheless, this interpretation is not sufficient to well explain the changes we observed, even if we assume that all the quantity of hydrogen incorporated in the material during deposition has out diffused. Indeed, after an annealing at 600 °C, $E_g$ and $E_u$ values are very different from those measured on the sample which was deposited in the same experimental conditions, under no hydrogen pressure (figure 5). This proves that, even if the out diffusion of hydrogen is total, it cannot explain only by itself the observed results. Taking in consideration the values of $E_g$ and $E_u$ that we found, the most plausible interpretation of our results should combine hydrogen out diffusion to an efficient process of crystallisation of the layers, which is well favoured by the high values of $T_R$ [33, 37, 40, 45, 46]. However, the creation of new structural defects in the material, due to thermal constraints and heating, is not excluded [47].

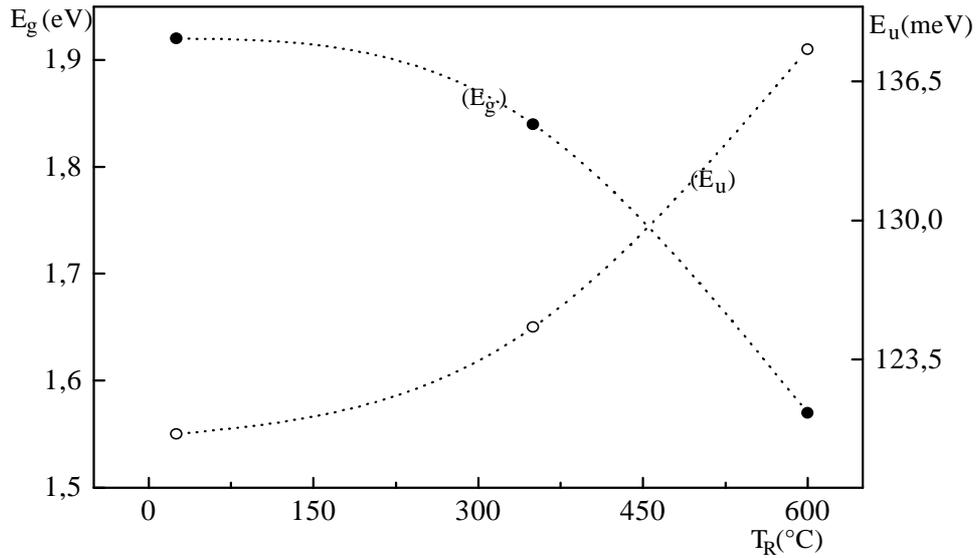

**FIGURE 10**

*Variations of the optical gap $E_g$ and the Urbach energy $E_u$ as function of the annealing temperature $T_R$. The strong co-relation linking these two parameters makes their curves completely symmetrical.*

X-rays diffraction measurements performed on thermally annealed thin films confirmed that no easily observable micro-crystallinities occur in the layers [37, 38]. From grazing x-rays reflectometry analysis, more interesting results have been obtained [20]. On figure 11, we have reproduced in semi-logarithmic scale a typical experimental curve plotting the variation of the reflected intensity I with the incidence angle α after a post-deposition annealing at a temperature of 500°C for 30 min. The linear fitting of this curve is shown in Figure 12. The exploitation of the obtained straight line provides us with the experimental parameters presented in the last row of table I. The comparison of these values with those obtained on the not annealed similar sample (sample C) shows clearly an increase of ρ accompanied of a decrease of Δe. These modifications resulting from the thermal annealing treatment could be the result of a micro-scale atomic rearrangement due to a starting partial re-crystallisation process in the deposited layer. Indeed, many literature results show that amorphous silicon crystallises during a traditional thermal annealing at a critical temperature between 550 °C and 700 °C [33, 39]. As we mentioned before, the increase of ρ is explained by that observed on the refraction index n calculated from optical measurements using transmission spectra.

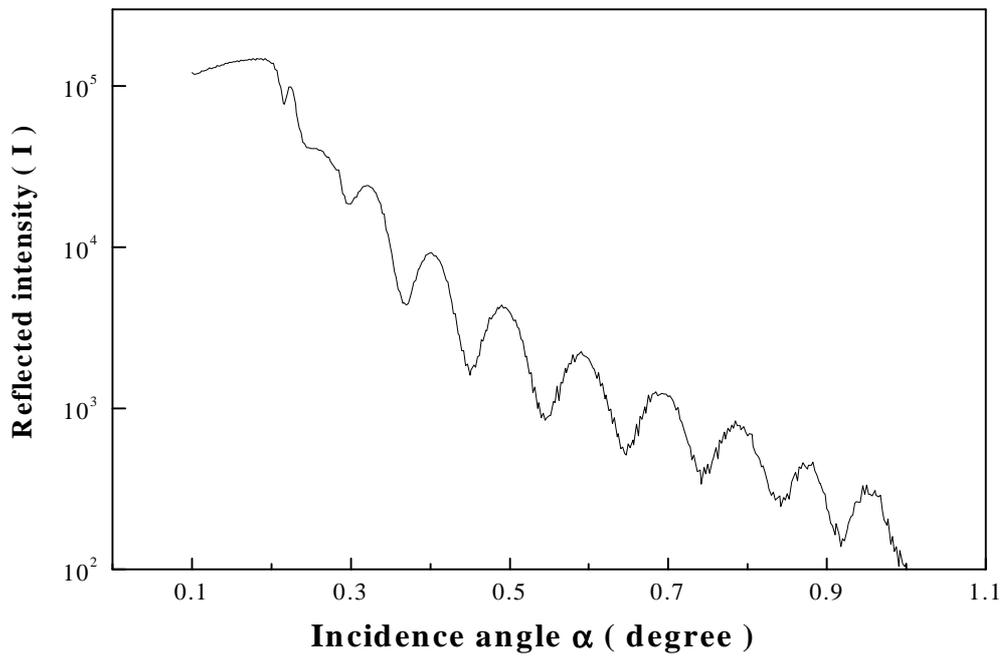

**FIGURE 11**
*Semi-logarithmic plotting of the reflected intensity I versus the incidence angle α for the sample D analysed after a post-deposition thermal annealing.*

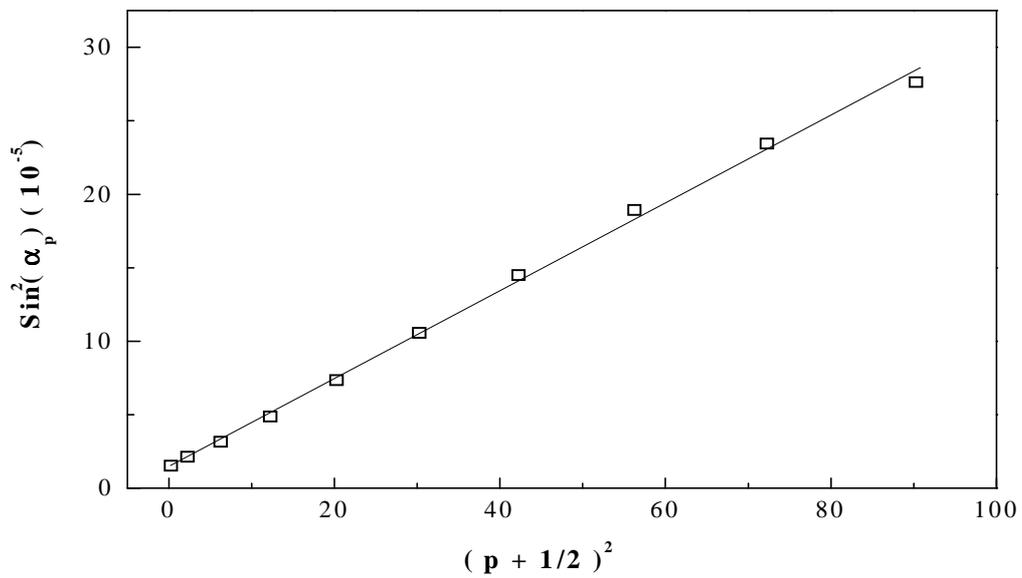

**FIGURE 12**
*Linear fitting of the experimental curve shown in figure 11*

On figure 13, we have presented in logarithm scale the variation of X-rays penetration depth $l$ in thin films as a function of the incidence angle $\alpha$ for samples A and D. We observe that for low $\alpha$ ($\alpha < 0.22°$), $l$ is slightly deeper in A (without hydrogen) than in the other sample (D). This is because the mass density $\rho$ of sample A is lower. For higher values of $\alpha$, $l$ increases first steeply at $\alpha_c$ and slowly to reach a similar level for both A and D.

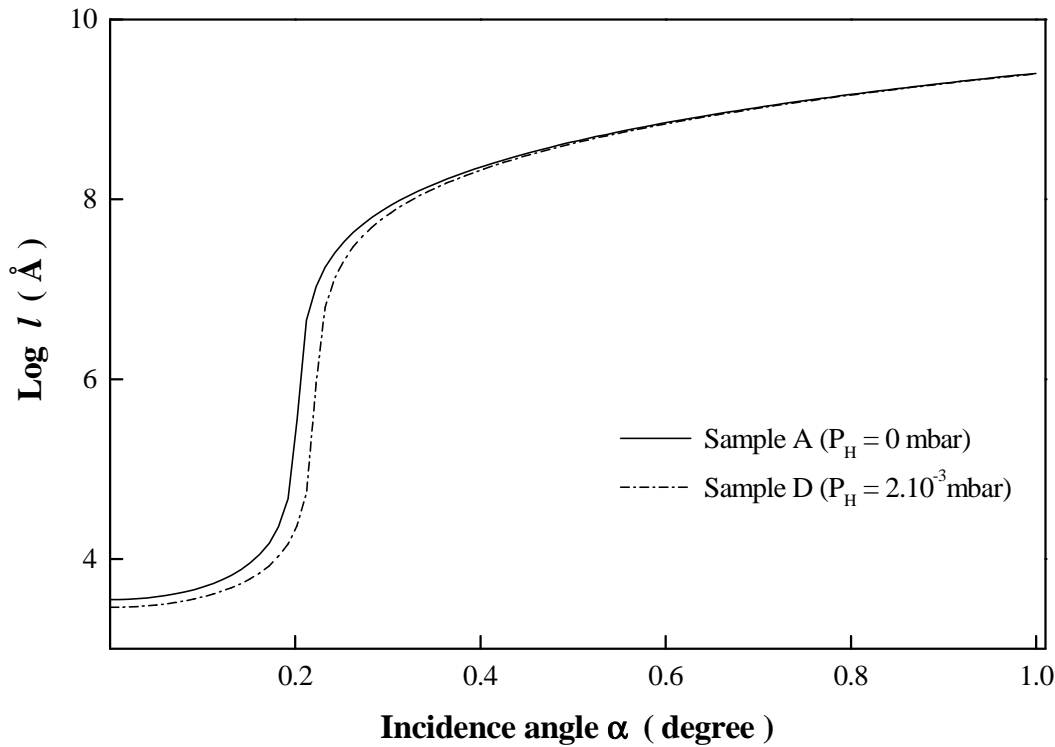

**FIGURE 13**

*Logarithmic variations of the penetration depth $\tau$ versus the incidence angle $\alpha$ in samples A deposited without measurable hydrogen partial pressure($P_H = 0$ mbar) compared to sample D deposited under hydrogen partial pressure $P_H = 2.10^{-3}$ mbar.*

### III -3- Measurements performed after surface oxidation

After thin films deposition and before taking reflectometry measurements, some samples named E, F and G are initially remained in the ambient for a long or short time without being directly exposed to solar radiations. $I(\alpha)$ curves taken on these samples are shown on figure 14. Curves a and b present a beat phenomenon. This means that the corresponding samples present an oxide coating on their surface. The oxide coating was to be naturally formed during the long stay of these samples in the ambient.

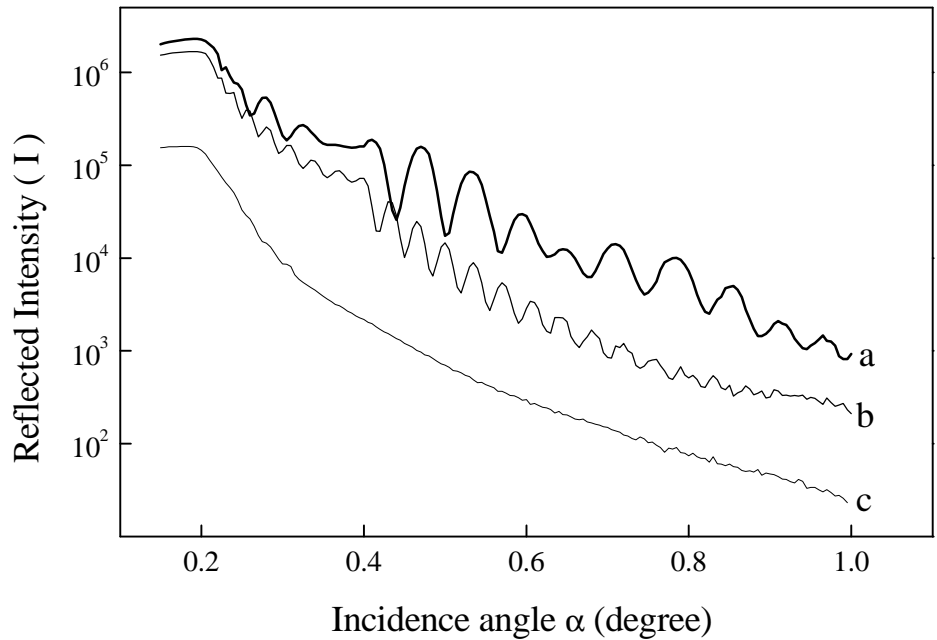

**FIGURE 14**

*Semi-logarithmic plotting of the reflected intensity I versus the incidence angle α measured on three a-Si:H thin films E, F and G after their surface oxidation in air ambient*

From the analysis of the above three experimental curves a, b and c, we have deduced the values of the characteristic parameters shown in table II. For samples E and F that remained in the ambient for the same period (40 days), we found that the thickness $\Delta e_{ox}$ and the electronic density $\rho_{ox}$ of the oxide coating are larger in the sample E which was deposited without any hydrogen incorporation. This could be explained by the fact that hydrogen existing in sample F reduces the density of defects acting as nucleation centres for oxide growth [48]. Moreover, during the deposition when the partial pressure of hydrogen increases, the density of Si-Si bonds decreases while that of Si-H bonds, which are energetically more stable [49], increases. So, the more the sample is hydrogenated, the more its oxidation is slow.

| Samples | Hydrogen partial pressure $P_H$ ( mbar ) | Total thickness $\Delta e_{tot}$ ( Å ) | Oxide layer Thickness $\Delta e_{ox}$ ( Å ) | Silicon layer Thickness $\Delta e_{Si}$ ( Å ) | Mass density of thin oxyde layer $\rho_{ox}$ ( g . cm$^{-3}$ ) |
|---|---|---|---|---|---|
| E | 0 | 1257 | 952 | 305 | 2.1865 |
| F | $5.10^{-4}$ | 812 | 660 | 152 | 2.1318 |
| G | $2.10^{-3}$ | >2000 Å | - | - | 2.1906 |

**TABLE II**

*Parameters values determined from X-rays analyses performed on the samples E, F and G.*

Curve c giving I($\alpha$) for the sample G does not show Kiessing fringes. Since this sample was remained in the ambient for a very long time (six months), it should be oxidised so much that its thickness exceeded the typical value of 2000 Å, which limits our calculation validity.

Finally, it will be interesting to notify that the density of oxide formed on the surface of our samples is lower than that of silice ($SiO_2$). This indicates that the formed oxide layer contains a high density of vacancies and thus it is quite fragile.

## IV- CONCLUSION

In the present work, we have studied the effect of partial hydrogen pressure $P_H$, and that of the classical annealing on the optical characteristics of a-Si:H thin films prepared by radio frequency cathodic sputtering. The obtained results show that each of the optical gap $E_g$ and the Urbach energy $E_u$ allow a very good control of the optical characteristics of the deposited layers. For low $P_H$, hydrogen saturates most of dangling bonds existing in these layers. This is shown through the clear improvement of $E_g$ and $E_u$. This improvement itself is optimal when $P_H = 3.10^{-3}$ mbar. For higher $P_H$, hydrogen begins to create its own defects and the optical performances of the layers start decreasing. After a first annealing at 350 °C, a high quantity of hydrogen incorporated in the layers out-diffuses. After a second annealing at 600 °C, hydrogen out diffusion is higher but it is accompanied by partial re-crystallisation of the layers. This is why structural quality of the latter shows a clear improvement. Using the grazing X-rays reflectivity technique we have also characterise our a-Si:H thin films immediately after deposition as well as after a post-deposition thermal annealing or natural oxidation in the ambient. Analysis of experimental curves obtained from examined samples enabled us to determine some physical parameters, particularly the thickness and mass density of deposited layers. We show that, when the partial pressure of hydrogen during the deposition process raise, the density of deposited layers increases and their thickness decreases. These results are explained by the fact that increasing the hydrogen concentration in the deposition reactive plasma leads to both layers densification mechanism, resulting from dangling bonds saturation, and a deposition rate reduction. We show also that hydrogen plays a protective role against surfaces oxidation of deposited layers staying in the ambient for a short time. This role disappears when the stay duration in the ambient is so long. Owing to the fact that X-rays reflectivity technique works in the ambient, samples studied immediately after their deposition should normally present a very thin natural oxide layer on their free surfaces. However, the thickness of such oxide layer is so thinner that it has no negative effect on the calculation method we have used. In our progressing work, we are trying to determine experimentally the thickness of this layer.

Currently, a new a-Si:H thin films deposition set up based on Hg-sensitized photo-CVD [50] technique is starting to develop in home laboratory. The research program that we will carry out on this new system registers within the framework of a research project financed by our ministry of high education and scientific research. It mainly consists in deposing a good quality a-Si:H thin films allowing to developing powerful terrestrial solar cells.

**ACKOWLEDGMENTS**

This paper has been prepared on the basis of experimental results obtained at home laboratory in morocco. It has been structured, enriched by recent literature data and written during the visit of the first author A. Barhdadi as Regular Associate Professor at the Abdus Salam International Centre for Theoretical Physics (ICTP), This author would like to thank the Director and staff of the Centre for the generous hospitality, the great support, and the efficient assistance. He also wishes to thank Professor G. Furlan, Head of ICTP-TRIL Programme, for his kind scientific co-operation and paper referring. The authors thanks are also addressed to Professors E. L. Ameziane and M. Azizane from "Physique du Solide" Laboratory, Faculty of Sciences Semlalia, Marrakech (Morocco), for their kind and precious scientific co-operation and efficient technical help.